\documentclass[10pt,
							 twocolumn,
							 superscriptaddress,
							 english,
							 prl,
							 showpacs,
							 floatfix,
							 aps
							]{revtex4-1}

\usepackage[utf8]{inputenc}
\usepackage{amsmath}
\usepackage{amssymb}
\usepackage{graphicx}
\usepackage{xspace}

\makeatletter

\@ifundefined{textcolor}{}
{%
 \definecolor{BLACK}{gray}{0}
 \definecolor{WHITE}{gray}{1}
 \definecolor{RED}{rgb}{1,0,0}
 \definecolor{GREEN}{rgb}{0,1,0}
 \definecolor{BLUE}{rgb}{0,0,1}
 \definecolor{CYAN}{cmyk}{1,0,0,0}
 \definecolor{MAGENTA}{cmyk}{0,1,0,0}
 \definecolor{YELLOW}{cmyk}{0,0,1,0}
}

\usepackage{soul}
\usepackage{braket}
\usepackage[backref=none,
bookmarksnumbered=true,
bookmarks=true,
bookmarksopen=true,
colorlinks=true,
citecolor=blue,
linkcolor=blue,
anchorcolor=green,
urlcolor=blue,unicode=false]{hyperref}

\let\baraccent=\= 
\renewcommand{\=}[1]{\stackrel{#1}{=}} 

\newcommand{\didv}{\ensuremath{\mathrm{d}I/\mathrm{d}V}\xspace}

\makeatother

\usepackage{babel}

\def\didv{d$I$/d$V$}

\begin{document}

\title{Correlation of Kondo effect and molecular conformation of the acceptor molecule in the TTF-TCNE charge transfer complex}

\author{Paul Stoll}
\affiliation{\mbox{Fachbereich Physik, Freie Universit\"at Berlin, Arnimallee 14, 14195 Berlin, Germany}}

\author{Christian Lotze}
\affiliation{\mbox{Fachbereich Physik, Freie Universit\"at Berlin, Arnimallee 14, 14195 Berlin, Germany}}

\author{Janina N. Ladenthin}
\affiliation{\mbox{Fachbereich Physik, Freie Universit\"at Berlin, Arnimallee 14, 14195 Berlin, Germany}}

\author{Tobias R. Umbach}
\affiliation{\mbox{Fachbereich Physik, Freie Universit\"at Berlin, Arnimallee 14, 14195 Berlin, Germany}}

\author{Isabel Fern\'andez-Torrente}
\affiliation{\mbox{Fachbereich Physik, Freie Universit\"at Berlin, Arnimallee 14, 14195 Berlin, Germany}}

\author{Katharina J. Franke}
\affiliation{\mbox{Fachbereich Physik, Freie Universit\"at Berlin, Arnimallee 14, 14195 Berlin, Germany}}


\date{\today}

\begin{abstract}
A Kondo resonance has been observed on purely organic molecules in several combinations of charge transfer complexes on a metal surface. It has been regarded as a fingerprint of the transfer of one electron from the donor to the extended $\pi$ orbital of the acceptor's LUMO. Here, we investigate the stoichiometric checkerboard structure of tetrathiafulvalene (TTF) and tetracyanoethylene (TCNE) on a Au(111) surface using scanning tunneling and atomic force microscopy at $4.8$\,K. We find a bistable state of the TCNE molecules with distinct structural and electronic properties. The two states represent different conformations of the TCNE within the structure. One of them exhibits a Kondo resonance, whereas the other one does not, despite of both TCNE types being singly charged.

\end{abstract}

%
%
%
%
%

\maketitle 

\section{Introduction}

Organic charge transfer complexes are crystalline materials composed of an electron donating and electron accepting molecular species. Typically, less than one electron is transferred from the donor to the acceptor~\cite{FraxedasBook}. Deposition of these molecules on metal surfaces results in various self-assembled structures with the amount of charge transfer depending on the exact composition~\cite{Jackel2008, Fiedler2013, Rodriguez2017}. Interestingly, several 1:1 stoichiometric structures feature a charge transfer of one electron into the acceptor's LUMO~\cite{Fernandez-Torrente2008, Torrente2012, Umbach2013, Wackerlin2011}. This unpaired electron spin yields intriguing magnetic phenomena, such as the emergence of a Kondo resonance~\cite{Fernandez-Torrente2008, Torrente2012, Umbach2013}. 

The Kondo resonance can be regarded as a useful tool to verify a singly charged molecular state. On the contrary, in absence of a Kondo resonance, the charge state cannot be identified. One should note that the absence also does not negate the single occupation of a molecular state, because other influences, such as the exchange coupling strength to the substrate may obscure the detection of a Kondo resonance at finite temperature. 
Indeed, it has been shown that slight modifications of the chemical structure or intramolecular conformations of individual molecules with an incorporated d-metal ion can lead to significant changes in the Kondo temperature~\cite{Zhao2005,Iancu2006}.

Here, we employ a combination of Kondo spectroscopy and measurements of the local contact potential difference (LCPD) to identify the charge state of TCNE molecules within the charge transfer complex of TTF-TCNE on a Au(111) surface. We find that the TCNE molecules adopt a metastable configuration in these monolayer islands. One of the species exhibits a Kondo resonance. LCPD values reveal that both molecules carry a similar charge. AFM images recorded with a functionalized tip in the repulsive regime allow for the resolution of a rather straight and a bent TCNE configuration, respectively. We conclude that the conformation determines the exchange coupling strength with the conduction electrons of the substrate giving rise to significantly different strengths of the Kondo screening.

\section{Experimental Details}

A Au(111) single crystal was cleaned by standard cycles of Ne$^{+}$ sputtering and annealing under ultra-high vacuum. Tetrathiafulvalene (TTF) and Tetracyanoethylene (TCNE) were placed in pumped sealed glass tubes connected to the preparation chamber. The molecules were sublimated by heating the glass tubes to $315-320$\,K and dosed through leak valves onto the Au(111) sample held at $<220$\,K. The sample was then annealed to $300$\,K to obtain highly ordered monolayer islands. This temperature is above the evaporation
temperature of TCNE from the Au(111) surface but below the evaporation temperature of TCNE from the mixed TTF-TCNE islands.
The as-prepared sample was pre-cooled and transferred into an STM equipped with a qPlus tuning fork sensor. All data was recoded at a temperature of $4.8 $\,K. Differential conductance spectra (\didv) were acquired with a standard lock-in technique in open feedback loop conditions.

\section{Results and Discussion}

\subsection{Self-assembly of TTF-TCNE on Au(111)}

\begin{figure*}[tb!]
\begin{center} 
\includegraphics[width=12 cm]{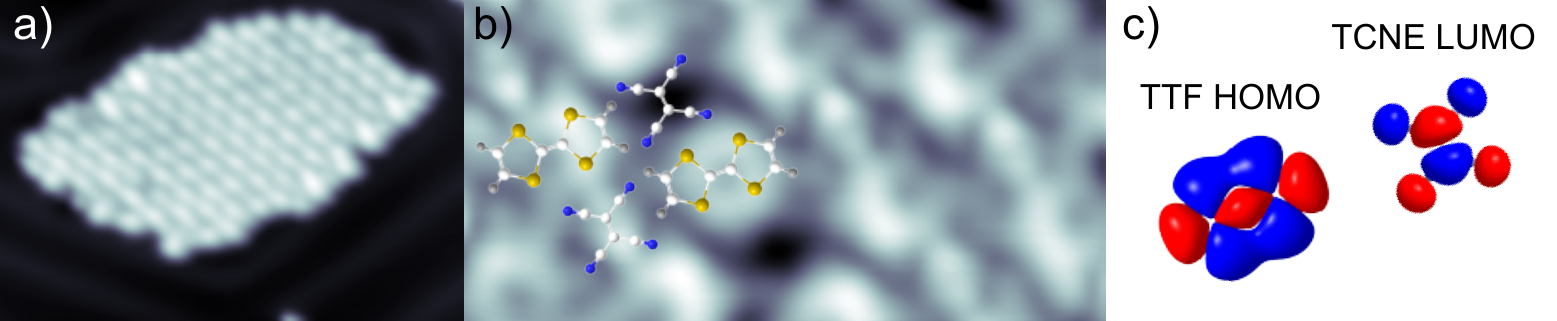}
 \caption{Constant-current STM image of a) a self-assembled TTF-TCNE island on a Au(111) surface. The most abundant structure is a 1:1 checkerboard arrangement. b) Close-up view of 1:1 checkerboard structure with superimposed structural models of individual molecules (scanning parameters $V=12$\,mV, $I=0.2$\,nA). The seemingly missing TCNE molecule is in the low conducting state. c) Calculated shape of the HOMO of the donor TTF and LUMO of the acceptor TCNE using Gaussian\;09 package, employing the 6-31G+dp basis set.} 
\label{fig1}
\end{center}

 \end{figure*}

\begin{figure*}[tb!]
\begin{center} 
\includegraphics[width=12 cm]{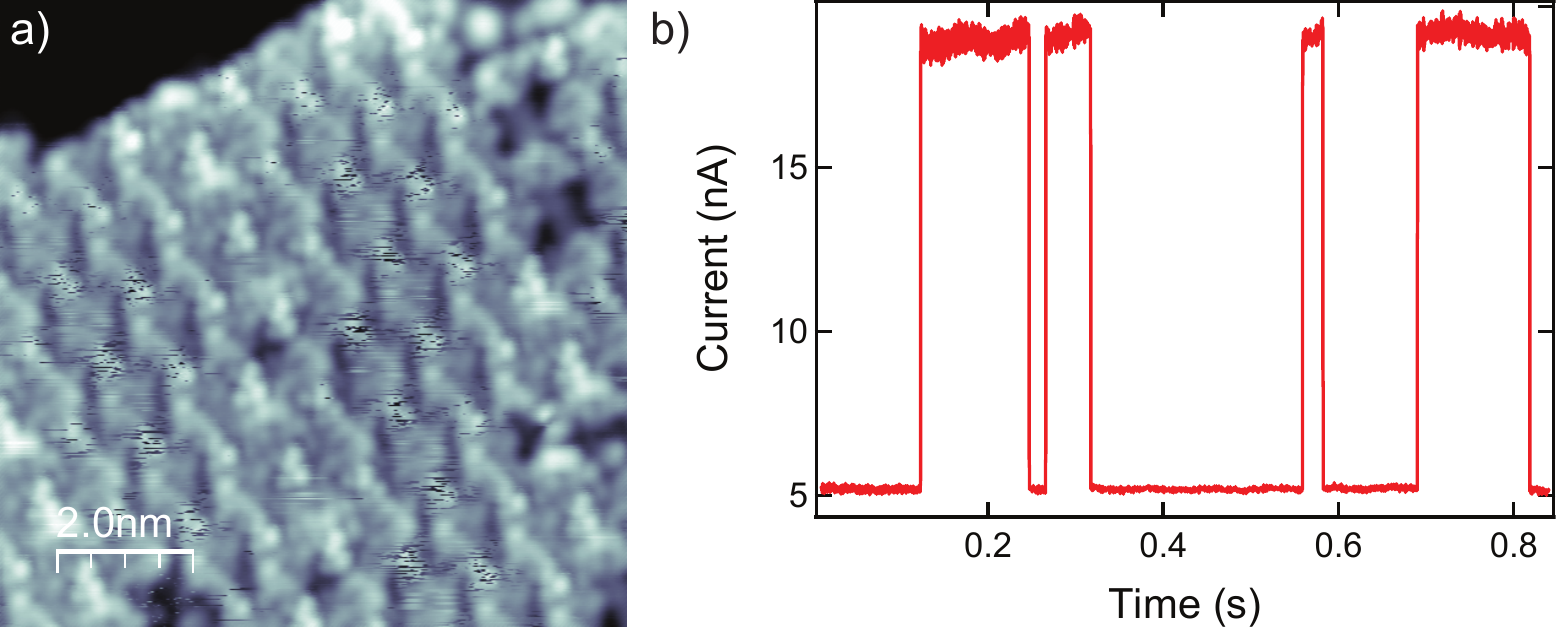}
 \caption{a) Constant-current STM image recorded at $V=-350$\,mV, $I=0.3$\,nA of the self-assembled checkerboard structure of TTF-TCNE showing a flickering of TCNE molecules. b) The bistability is reflected in two-level fluctuations of the tunneling current on an individual TCNE molecule at $V=-150$\,mV.} 
\label{fig2}
\end{center}
\end{figure*}

STM images of the as-prepared sample reveal highly-ordered islands of the TTF and TCNE molecules with the most abundant structure being a checkerboard pattern (Fig.~\ref{fig1}a). At low bias voltages, the TTF molecules can be recognized by their doughnut-like appearance, which resembles their HOMO (Fig.~\ref{fig1}b,c). In between we find tiles of "bright" and "dark" contrast. The "bright" tiles reveal an elongated protrusion with a slightly lower contrast in the center. This appearance agrees with the LUMO shape of the free TCNE molecule (Fig.~\ref{fig1}c), which exhibits electron density around the  C=C double bond with a nodal plane perpendicular to this axis. The cyano moieties are expected to also carry large electron density. Inspection of the STM images suggests that these lie close to the TTF molecules. 
At first sight, one may be tempted to assign the "dark" tiles to an empty space between the molecules. However, scanning at larger bias voltages shows a flickering noise at these positions (see Fig.~\ref{fig2}a). Recording a current-time ($I-t$) trace on these positions (see Fig.~\ref{fig2}b) reflects the presence of two distinct conductance states. These conductance levels represent the "bright" and "dark" TCNE molecules in Fig.~\ref{fig1}b). The molecules may end up stochastically in the "bright" or "dark" state when the bias voltage and tunneling current are reduced.

We conclude that the overall structure of the island is a 1:1 phase of TTF and TCNE with the molecules lying almost parallel to the surface. A small tilt of the TTF molecules can indeed be inferred from the slightly asymmetric doughnut shape of TTF. Similar tilt angles have been observed for individual TTF molecules on Au(111) as a result of covalent bonds of the S atoms with the underlying substrate~\cite{Fernandez-Torrente2007}. The configuration of the TCNE molecules cannot be easily inferred from the STM images, neither in the high-conductance nor in the low-conductance state. It is also not \textit{a priori} clear whether the two distinct states are different conformational states, where the topography dominates the appearance in the STM images, or different charge states, whose electronic properties determine the appearance in the STM images.

\subsection{Kondo effect on bistable TCNE conformations}

\begin{figure*}[tb!]
\begin{center} 
\includegraphics[width=12 cm]{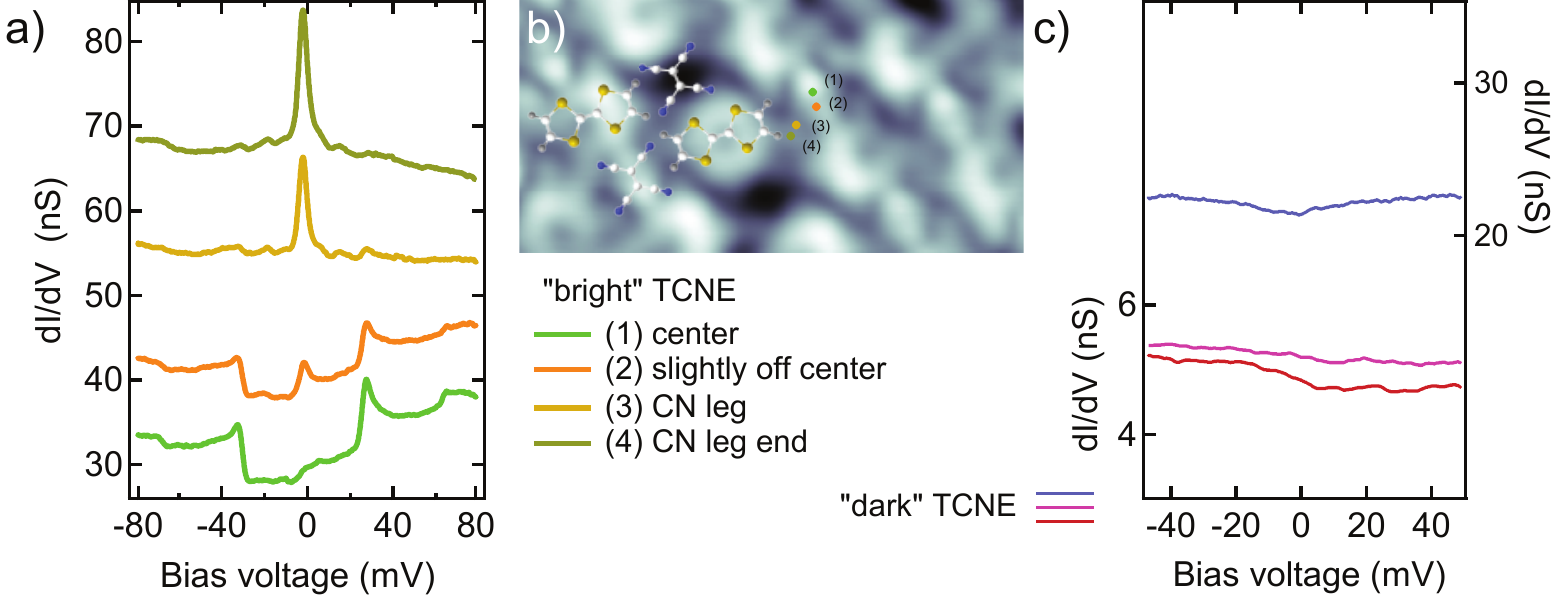}
 \caption{a) \didv\ spectra taken on different locations of a "bright" TCNE molecule, which are indicated in the STM image in b). The spectra are vertically offset by 10\,nS each for clarity. Feedback was opened at $V=80$\,mV, $I=3$\,nA. c) \didv\ spectra on "dark" TCNE measured with three different tips on different molecules. Feedback was opened at $V=100$\,mV, $I=0.5$\,nA (red, purple) and at $V=50$\,mV, $I=1$\,nA (blue). } 
\label{fig3}
\end{center}
 \end{figure*}

To unravel the nature of the two states, we record \didv\ spectra on the TCNE molecules. The "bright" TCNE molecules exhibit a narrow peak at the Fermi level on the cyano moieties (Fig.~\ref{fig3}a). This peak is reduced in intensity on the center of the molecule. Instead, peaks on top of steps are located symmetrically around the Fermi level at $\pm 30$\,mV. These spectra are reminiscent of \didv\ spectra on tetracyanoquinodimethane (TCNQ) molecules mixed with TTF or tetramethyl-TTF on Au(111)~\cite{Fernandez-Torrente2008, Torrente2012}. The narrow zero-bias anomaly has been identified as a Kondo resonance, whose width is determined by the strength of the exchange scattering of the conduction electrons on the unpaired electron spin in the $\pi$ orbital. The large amplitude of the Kondo resonance at the cyano moiety reflects a large electron density at these terminations due to their electrophilic character~\cite{Umbach2013}. While the Kondo resonance is almost symmetric at the cyano terminations, the slight asymmetry in the TCNE center may be ascribed to interference between the tunneling paths into the Kondo resonance and the substrate~\cite{Ternes2009}. The presence of the Kondo resonance thus signifies an electron being located in the LUMO due to the charge transfer from the environment. 
The steps at $\pm 30 $\,mV arise from inelastic vibrational excitations~\cite{Stipe1998a, Wegner2009}. The threshold energy matches the out-of-plane rocking mode or wagging mode of the TCNE molecule~\cite{Michaelian1982}. Density functional theory calculations for TCNE molecules on Ag(100) showed that these steps correspond to the rocking mode~\cite{Wegner2013}.
 The largest step height is naturally found at the center, where the rocking mode most strongly affects the conductance. The convolution with a peak has previously been ascribed to the interplay of the Kondo effect with molecular vibrations~\cite{Fernandez-Torrente2008, Paaske2005, Iancu2016}. 
The \didv\ spectra on the "dark" TCNE molecules are essentially flat (Fig.~\ref{fig3}c). In particular, there is no indication of a Kondo resonance. 
It is worthy to note that TCNE molecules alone on a Au(111) surface do not exhibit a Kondo resonance~\cite{Wegner2008}. Hence, one may be tempted to use the presence/absence of a Kondo resonance to determine the charge state of a molecule. This would indicate that the "dark" molecule is not singly charged in contrast to the "bright" state. To verify or falsify this assumption, we need to employ another method for the charge-state determination of individual molecules. 
 
\subsection{Charge state of TCNE molecules}

Recently, non-contact atomic force microscopy (nc-AFM) has been successfully used to determine charge states of single atoms and molecules on surfaces~\cite{Gross2009a,Mohn2012}. The principle is based on the measurement of the local contact potential difference (LCPD), which is caused by the electric field of localized charges at tip and sample. 

\begin{figure*}[tb!]
\begin{center} 
\includegraphics[width=12 cm]{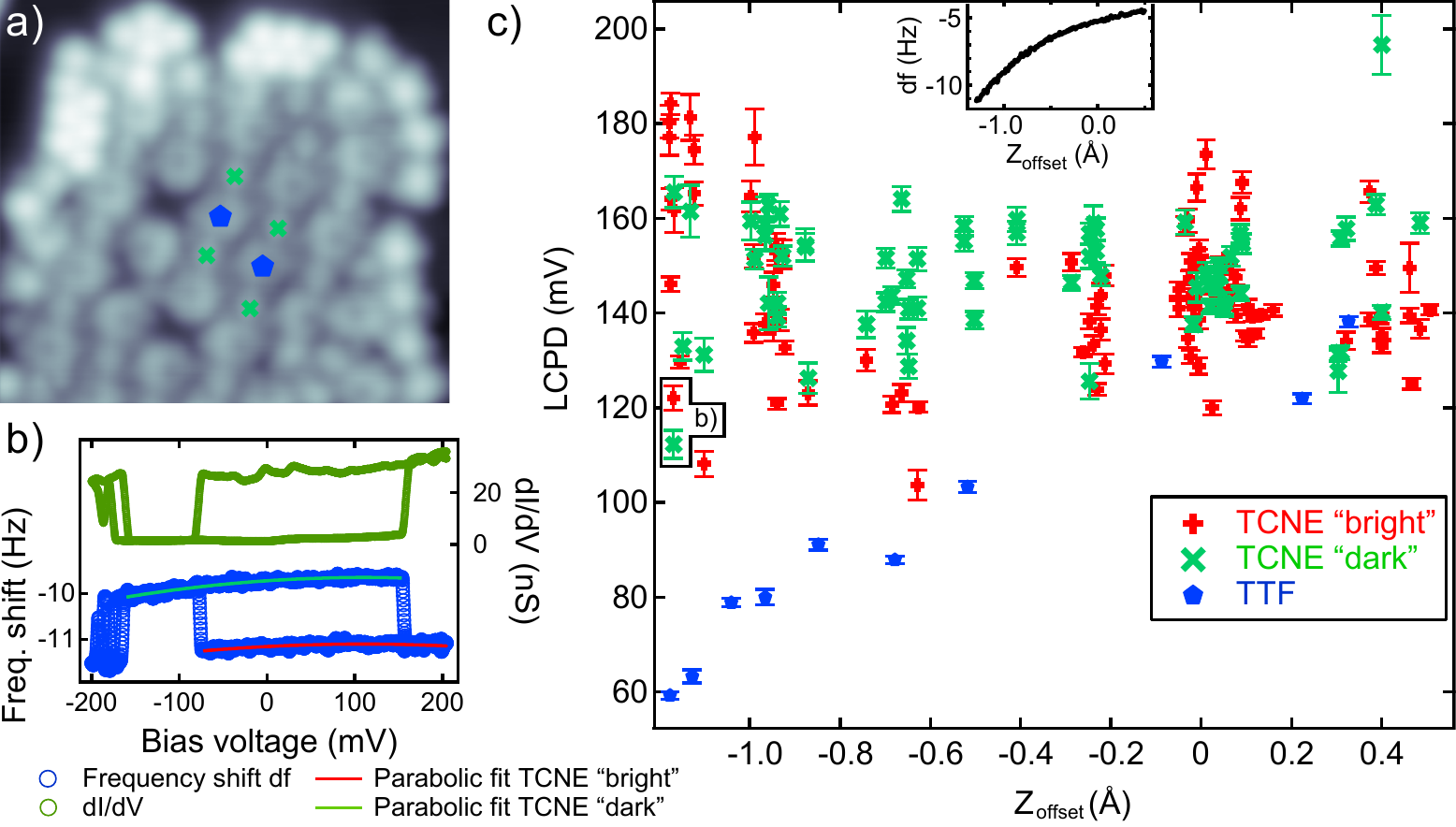}
 \caption{a) STM topography of a TTF-TCNE island with some Xe atoms at the island's edge. The Xe atoms have been used for tip functionalization. b) Exemplary $df-V$  and \didv\ curves at low bias voltage of a TCNE molecule in its two conformational states. The $df-V$ curves were fitted by a parabola with its vertex determining the LCPD value. c) Extracted LCPD values for two TTF molecules (blue pentagons in a) and four TCNE molecules (green crosses in a) as a function of tip--molecule distance. $Z_\mathrm{offset}$ determines the topographic height above an STM setpoint of $V=80$\,mV, $I=60$\,pA on a "bright" molecule. The inset shows a $df-Z$ curve illustrating that all LCPD data was taken in the attractive regime.} 
\label{fig4}
\end{center}
 \end{figure*}

We employ a Xenon-functionalized tip, which provides a chemically inert tip apex and allows to approach the molecule down to the repulsive regime without affecting the molecular integrity~\cite{Gross2009b}. As the local environment may influence the LCPD signal, we compare several TTF and TCNE molecules. The most efficient strategy to minimize the effect of changes in the local environment is to measure the LCPD on the very same TCNE molecule, both in its "bright" and "dark" state. Figure~\ref{fig4}b shows the $I-V$ signal taken in a backward and forward sweep in a voltage range of $\pm 200$\,mV. The two different slopes reflect the two different states of higher and lower conductance of the "bright" and "dark" TCNE molecule, respectively. The switching between the two states is seen as a sudden drop/increase in the current. The simultaneously recorded $df$ signal is displayed in the same figure. Both states exhibit a distinct $df-V$ signal that can be fit by a parabolic shape. The LCPD value can be extracted from the vertex of the parabolas (even though the maximum may lie outside of the measurable voltage range). The values shown here are all determined from $df-V$ curves, which are recorded in the attractive part of the interaction potential. Hence, we can exclude that repulsive forces may induce deformations in the tip or in the underlying molecule. \footnote{$df$ data was corrected for $Z$-creep by setting the initial and final point of the forward-backward spectrum to the same value. The deviation throughout the forward-backward scan was fitted by a linear and one exponentially decaying $Z$ value as correction factor. After this correction the largest error stems from the noise in the $df$ signal and the fact that the stable $df-V$ range is rather small.}

Figure~\ref{fig4}c shows a compilation of the LCPD values of TTF  and TCNE molecules (shown in Figure~\ref{fig4}a) in their "bright" and "dark" state. At large tip--sample distance, we cannot resolve any difference of LCPD values anywhere. With decreasing distance, the LCPD measured on the TTF molecules separates from the values on the TCNE molecules. A lower LCPD value signifies a more positive charge. The lower LCPD on TTF than on TCNE is thus in agreement with the expectations for charge transfer between the TTF, the TCNE and the Au(111) surface. The indistinguishability at large tip--sample distance can be ascribed to a loss of spatial resolution due to an increase of the area affecting the measured LCPD value. At very close distance, the extension of the molecular wave functions into vacuum has to be taken into account~\cite{Schuler2014}. Importantly, our plot suggests a clear distinction between the charge states of TTF and TCNE in a range of tip--sample distance. Interestingly, we do not find a sizable difference between the "bright" and "dark" state of the TCNE molecules. The LCPD values of the TCNE molecules scatter significantly around   $(140\pm 20)$\,meV. However, we note that the "bright" and "dark" state cannot be separated, but are equally scattered around the average value. If we compare the values of "bright" and "dark" state measured in a single bistable forward-backward $df-V$ scan (as in Fig.\ref{fig4}b), the difference between the two configurations amounts to only $(10\pm 15)$\,meV. Comparing this difference to the LCPD value of TTF being $(80\pm20)$\,meV lower allows to estimate an upper bound on the difference in the charge states of the two TCNE molecules. The TTF molecule is probably positively charged with - at maximum - a single positive charge. The presence of a Kondo resonance on the "bright" TCNE molecules signified a single negative charge state. 
Consequently, we can assume that a charge difference of $2 e$ corresponds to a difference in the LCPD value of $\sim 80$\,meV. With the difference between the TCNE states being $(10\pm 15)$\,meV, we arrive at a difference in the charge state of $(0.25\pm 0.38) e$ for "bright" and "dark" TCNE state.  We can therefore assume that both TCNE states exhibit roughly the same charge, \textit{i.e.}, are both carrying one electron in their LUMO. 
This seems to be a surprising result in view of the absence of a Kondo resonance on the "dark" TCNE molecule.

\subsection{Molecular conformations of the "bright" and "dark" TCNE molecules}

In the next step, we aim at resolving the difference between the "bright" and "dark" molecular states, which leads to the presence or absence of a Kondo resonance despite of similar charging. The observation of a Kondo resonance does not only require the presence of an unpaired electron spin but also sufficient exchange coupling with the substrate electrons~\cite{Ternes2009}. The strength of the exchange coupling is given by the hybridization of the spin-carrying molecular orbital and the electronic states of the substrate. Hence, changes in the adsorption height and/or conformational changes may play a decisive role. 
To gain further insights into the adsorption geometry, we use frequency-shift imaging and spectroscopy of our nc-AFM, again with a Xenon-functionalized tip apex.

\begin{figure*}[tb!]
\begin{center} 
\includegraphics[width=12 cm]{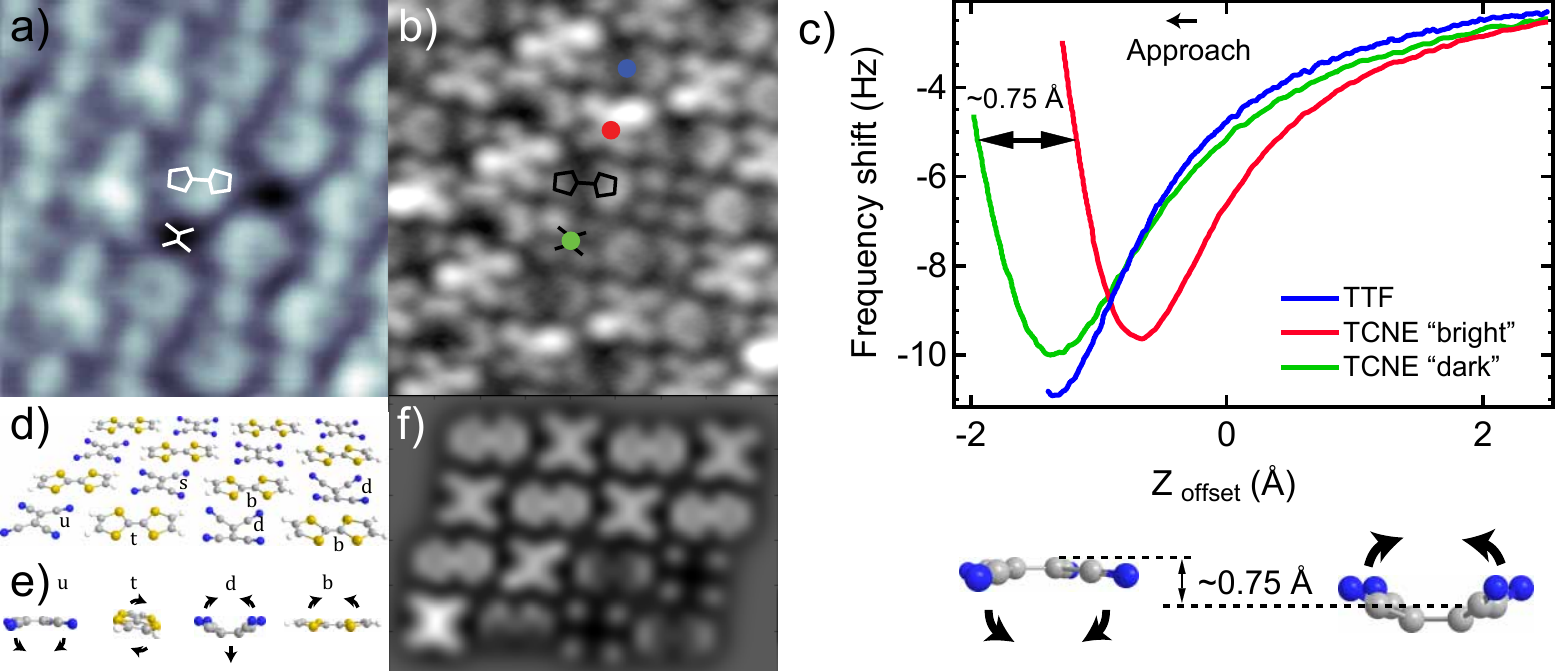}
 \caption{a) Constant-current STM image of a TTF-TCNE island ($V=90$\,mV, $I=57$\,pA). b) $df$ image recorded at a constant height given by a plane fixed at $V=90$\,mV, $I=57$\,pA STM feedback on a "dark" TCNE and approaching by $2$\,$\mathrm{\AA}$. c) $df-Z$ curve taken on TTF and TCNE molecules with their positions indicated in color in b). $Z_\mathrm{offset}$ is measured from the same plane as in b). The repulsive regime is shifted by $\sim 0.75 $\,$\mathrm{\AA}$ between the "dark" and "bright" TCNE, suggesting the bending of the TCNE molecules as indicated in the sketch.  d) $4\times 4$ checkerboard model structure of a TTF-TCNE island arrangement. The upper two rows consist of flat molecules, the lower two rows consist of differently deformed molecules. $s$ marks a straight TCNE molecule, $b$ a bent TTF molecule, $u$ up-tilted TCNE, $d$ down-tilted TCNE molecule, $t$ a tilted TTF molecule. e) Side view of the deformations in the lowest row of d). f) Simulated AFM image of the structure in d). }
\label{fig5}
\end{center} 
 \end{figure*}

Figure~\ref{fig5}b shows a constant-height frequency-shift image of a TTF-TCNE island, which was recorded at the same area as the STM image in Fig.~\ref{fig5}a. The TTF molecules appear with two tilted ring shapes (see details below). They are topographically lower than the TCNE molecules. The "bright" TCNE molecules appear as an elevated X shape, while the "dark" TCNE molecules appear as a depression with four protrusions at the place of the cyano terminations. These differences hint at large structural differences of the two TCNE species. More quantitative insights can be gained form $df-z$ curves taken at different sites on the molecules in 
 Fig.~\ref{fig5}c. All curves show the qualitative behavior of a Lennard-Jones potential with a long-ranged attractive part stemming from electrostatic forces and a short-ranged repulsive part originating from Pauli repulsion of the Xenon atom at the tip and the molecules~\cite{Moll2010}. The repulsive part of the curve on the X-shaped TCNE is shifted away from the surface by about $0.75$\,$\mathrm{\AA}$  as compared to the repulsive part on the center of the four-lobe TCNE species. Hence, we can conclude that the C=C double bond is found at significantly different distances from the Au(111) surface. A tentative sketch of the two molecular conformations is shown in the inset of Fig.~\ref{fig5}c.

Hapala and co-workers have developed a code which can be used  to simulate $df$ images  of molecules on a surface in the repulsive regime ~\cite{HapalaCode}. Their model uses a force field between a flexible tip apex (here: Xenon) and the molecules~\cite{Hapala2014,Hapala2014b}. We use their code and simulate the checkerboard structure consisting of a trial structure of flat and bent  TCNE configurations of Fig~\ref{fig5}c together with flat and bent TTF molecules. This composite structure of a $4\times 4$ checkerboard is shown in Fig.~\ref{fig5}d,e together with the simulated image in Fig.~\ref{fig5}f. The upper two rows represent the flat molecules, the two lower rows consist of differently bent molecular configurations. We first concentrate on the appearance of the TCNE molecules. The flat configuration shows up as a flat X in the AFM images. This is almost in agreement with some of the TCNE molecules in the experimental structure. A better agreement can be obtained by small tilt and bending angles, but we note that the details of this type of TCNE molecules may vary from molecule to molecule. Importantly, we note that the four-lobe structure of the TCNE cannot be produced by slight TCNE deformations. Only rather strong bending angles as suggested already by the $df-z$ curves in Fig.\ref{fig5}c can explain the characteristic appearance. The C=C double bond of the "dark" TCNE molecules resides much closer to the surface than the cyano moieties, which are bent upwards from the surface. Such bending can be facilitated by a slight change of hybridization from $sp^2$ towards $sp^3$ due to a backbonding with Au $d$-orbitals. Charge backdonation from transition metals to the strong $\pi$-acceptor TCNE is known for V(TCNE)$_2$ films~\cite{Kortright2008} and TCNE on a Cu surface~\cite{Choi2010}. Bending of the endgroups away from the metal is known for the structurally similar ethylene in Zeise's salt~\cite{miessler2011inorganic}.

We also note in passing that the image of the flat TTF molecule nicely reflects its intramolecular structure. Comparison of simulated and experimental images suggests that TTF also adopts a bent configuration. It is best described by up-tilted dithiol-ylidene rings, but with variations in the degree of bending and tilting along its long or short molecular axis. The details of these adsorption configurations probably influence the stability and favored state of the neighboring TCNE molecules. 

\subsection{Correlation of structural and electronic properties }

Having resolved two distinct TCNE states in STM and AFM measurements, finally enables us to draw a complete picture on the adsorption properties. Summarizing our complementary results, we have seen that the "bright" TCNE molecules are flat or only slightly bent with the CN terminations facing the surface. They show a Kondo resonance as an expression of their singly-occupied LUMO. In contrast, the "dark" TCNE molecules exhibit a bent configuration with the C=C bond being closest to the surface. These molecules do not exhibit a Kondo resonance, despite of a similar charge state as the "bright" species.
Hence, our results suggest a correlation of the presence of a Kondo resonance with molecular conformation.

To understand why the strongly bent "dark" TCNE molecule does not exhibit a Kondo resonance in contrast to the flat "bright" one, we discuss the requirements for the appearance of a Kondo resonance within the Anderson impurity model. First, we need a singly occupied orbital at energy $\epsilon_\mathrm{S}$ below the Fermi level $E_\mathrm{F}$. Second, the Coulomb repulsion $U$ needs to be sufficiently large such that the doubly charged orbital at $\epsilon_\mathrm{D}$ lies above $E_\mathrm{F}$. Third, the broadening $\Delta$ of these energy levels needs to be much smaller than the Coulomb repulsion ($\Delta << U$) in order to assure that fluctuations between the levels are negligible. The width of the Kondo resonance $\Gamma$ and the Kondo temperature $T_\mathrm{K}$ are then given by~\cite{Ternes2009}:
\begin{equation}
\Gamma=k_\mathrm{B} T_\mathrm{K} \cong \sqrt{2\Delta \frac{U}{\pi}}\mathrm{exp}\left(-\frac{\pi}{2\Delta}\left(\left|\frac{1}{\epsilon_\mathrm{S}}\right|+\left|\frac{1}{\epsilon_\mathrm{S}+U}\right|\right)^{-1}\right). 
\label{eq}
\end{equation} 

We assume that the Coulomb repulsion $U$ is similar in both types of molecules due to the same chemical nature. The parameters which may vary in the two configurations are the energy level alignments of $\epsilon_\mathrm{S}$ and the hybridization $\Delta$. We first note that $\Delta$ may be larger in the "dark" bent configuration, where the C=C double bond is rather close to the surface. However, the LUMO features a nodal plane at this position, whereas the electron-rich cyano groups point upwards. Hence, any directional bonding and hybridization with the conduction electrons of the surface is expected to be small.
Furthermore, we consider that the  "dark" bent configuration is stabilized by a redistribution of charges among the molecular orbitals~\cite{Choi2010}. The unpaired electron spin density may thus be located in orbitals  $\epsilon_\mathrm{S}$ that are further away from $E_\mathrm{F}$. According to equation~\ref{eq}, a spin-polarized orbital that is too far below  $E_\mathrm{F}$ leads to a very small Kondo temperature, which is below our measurement temperature of $4.8 $\,K.

In contrast, the "bright" molecular species is rather flat. The unpaired electron can therefore reside in the LUMO, which lies closely below $E_\mathrm{F}$.  The corresponding Kondo temperature is thus above the experimental temperature. Consequently, a Kondo resonance emerges at the Fermi level in the \didv spectra.

\section{Conclusions and Outlook}
Using a combination of STM- and AFM-based techniques, we could show that the electron acceptor TCNE forms a checkerboard monolayer with the electron donor TTF on a Au(111) surface. All TCNE molecules lie in a singly-charged state by accepting one electron in their LUMO. Two configurations are of similar stability but separated by a potential barrier. This energy landscape allows to observe two distinct conformational states. One is almost flat with only slightly down-bent cyano groups, whereas the other one is more strongly bent with the cyano groups facing upwards. In spite of the same charge state, the molecules with the upward bent CN moieties do not exhibit a Kondo resonance at an experimental temperature of $4.8$\,K. We ascribe its absence to a weak interaction of spin-carrying orbitals with the substrate's conduction electrons. 
Our results reflect the limits of using the presence or absence of a Kondo resonance as a tool for charge-state determination. Instead, an intriguing interplay of molecular conformation and Kondo screening could be resolved. 
The bistability of the TCNE molecules in the TTF-TCNE assembly demands for further investigations of the switching properties. Being able to control the conformational state may provide routes for reading and writing the absence/presence of a Kondo state at the molecular scale. 

\acknowledgments
We thank the Deutsche Forschungsgemeinschaft for support through the collaborative research centre SFB 658 and the European Research Council for the Consolidator Grant "NanoSpin".


\bibliographystyle{apsrev4-1}


\begin{thebibliography}{31}%
\makeatletter
\providecommand \@ifxundefined [1]{%
 \@ifx{#1\undefined}
}%
\providecommand \@ifnum [1]{%
 \ifnum #1\expandafter \@firstoftwo
 \else \expandafter \@secondoftwo
 \fi
}%
\providecommand \@ifx [1]{%
 \ifx #1\expandafter \@firstoftwo
 \else \expandafter \@secondoftwo
 \fi
}%
\providecommand \natexlab [1]{#1}%
\providecommand \enquote  [1]{``#1''}%
\providecommand \bibnamefont  [1]{#1}%
\providecommand \bibfnamefont [1]{#1}%
\providecommand \citenamefont [1]{#1}%
\providecommand \href@noop [0]{\@secondoftwo}%
\providecommand \href [0]{\begingroup \@sanitize@url \@href}%
\providecommand \@href[1]{\@@startlink{#1}\@@href}%
\providecommand \@@href[1]{\endgroup#1\@@endlink}%
\providecommand \@sanitize@url [0]{\catcode `\\12\catcode `\$12\catcode
  `\&12\catcode `\#12\catcode `\^12\catcode `\_12\catcode `\%12\relax}%
\providecommand \@@startlink[1]{}%
\providecommand \@@endlink[0]{}%
\providecommand \url  [0]{\begingroup\@sanitize@url \@url }%
\providecommand \@url [1]{\endgroup\@href {#1}{\urlprefix }}%
\providecommand \urlprefix  [0]{URL }%
\providecommand \Eprint [0]{\href }%
\providecommand \doibase [0]{http://dx.doi.org/}%
\providecommand \selectlanguage [0]{\@gobble}%
\providecommand \bibinfo  [0]{\@secondoftwo}%
\providecommand \bibfield  [0]{\@secondoftwo}%
\providecommand \translation [1]{[#1]}%
\providecommand \BibitemOpen [0]{}%
\providecommand \bibitemStop [0]{}%
\providecommand \bibitemNoStop [0]{.\EOS\space}%
\providecommand \EOS [0]{\spacefactor3000\relax}%
\providecommand \BibitemShut  [1]{\csname bibitem#1\endcsname}%
\let\auto@bib@innerbib\@empty
\bibitem [{\citenamefont {Fraxedas}(2006)}]{FraxedasBook}%
  \BibitemOpen
  \bibfield  {author} {\bibinfo {author} {\bibfnamefont {J.}~\bibnamefont
  {Fraxedas}},\ }\href@noop {} {\emph {\bibinfo {title} {Molecular Organic
  Materials: From Molecules to Crystalline Solids}}}\ (\bibinfo  {publisher}
  {Cambridge University Press},\ \bibinfo {address} {Cambridge},\ \bibinfo
  {year} {2006})\BibitemShut {NoStop}%
\bibitem [{\citenamefont {J\"ackel}\ \emph {et~al.}(2008)\citenamefont
  {J\"ackel}, \citenamefont {Perera}, \citenamefont {Iancu}, \citenamefont
  {Braun}, \citenamefont {Koch}, \citenamefont {Rabe},\ and\ \citenamefont
  {Hla}}]{Jackel2008}%
  \BibitemOpen
  \bibfield  {author} {\bibinfo {author} {\bibfnamefont {F.}~\bibnamefont
  {J\"ackel}}, \bibinfo {author} {\bibfnamefont {U.~G.~E.}\ \bibnamefont
  {Perera}}, \bibinfo {author} {\bibfnamefont {V.}~\bibnamefont {Iancu}},
  \bibinfo {author} {\bibfnamefont {K.-F.}\ \bibnamefont {Braun}}, \bibinfo
  {author} {\bibfnamefont {N.}~\bibnamefont {Koch}}, \bibinfo {author}
  {\bibfnamefont {J.~P.}\ \bibnamefont {Rabe}}, \ and\ \bibinfo {author}
  {\bibfnamefont {S.-W.}\ \bibnamefont {Hla}},\ }\href {\doibase
  10.1103/PhysRevLett.100.126102} {\bibfield  {journal} {\bibinfo  {journal}
  {Phys. Rev. Lett.}\ }\textbf {\bibinfo {volume} {100}},\ \bibinfo {pages}
  {126102} (\bibinfo {year} {2008})}\BibitemShut {NoStop}%
\bibitem [{\citenamefont {Fiedler}\ \emph {et~al.}(2014)\citenamefont
  {Fiedler}, \citenamefont {Reckien}, \citenamefont {Bredow}, \citenamefont
  {Beck},\ and\ \citenamefont {Sokolowski}}]{Fiedler2013}%
  \BibitemOpen
  \bibfield  {author} {\bibinfo {author} {\bibfnamefont {B.}~\bibnamefont
  {Fiedler}}, \bibinfo {author} {\bibfnamefont {W.}~\bibnamefont {Reckien}},
  \bibinfo {author} {\bibfnamefont {T.}~\bibnamefont {Bredow}}, \bibinfo
  {author} {\bibfnamefont {J.}~\bibnamefont {Beck}}, \ and\ \bibinfo {author}
  {\bibfnamefont {M.}~\bibnamefont {Sokolowski}},\ }\href {\doibase
  10.1021/jp407579z} {\bibfield  {journal} {\bibinfo  {journal} {J. Phys. Chem.
  C}\ }\textbf {\bibinfo {volume} {118}},\ \bibinfo {pages} {3035} (\bibinfo
  {year} {2014})} 
	 \BibitemShut {NoStop}%
\bibitem [{\citenamefont {Rodr\'iguez-Fern\'andez}\ \emph
  {et~al.}(2017)\citenamefont {Rodr\'iguez-Fern\'andez}, \citenamefont
  {Robledo}, \citenamefont {Lauwaet}, \citenamefont {Martín-Jim\'enez},
  \citenamefont {Cirera}, \citenamefont {Calleja}, \citenamefont
  {D\'iaz-Tendero}, \citenamefont {Alcam\'i}, \citenamefont {Floreano},
  \citenamefont {Dom\'inguez-Rivera}, \citenamefont {V\'azquez~de Parga},
  \citenamefont {\'Ecija}, \citenamefont {Gallego}, \citenamefont {Miranda},
  \citenamefont {Martín},\ and\ \citenamefont {Otero}}]{Rodriguez2017}%
  \BibitemOpen
  \bibfield  {author} {\bibinfo {author} {\bibfnamefont {J.}~\bibnamefont
  {Rodr\'iguez-Fern\'andez}}, \bibinfo {author} {\bibfnamefont
  {M.}~\bibnamefont {Robledo}}, \bibinfo {author} {\bibfnamefont
  {K.}~\bibnamefont {Lauwaet}}, \bibinfo {author} {\bibfnamefont
  {A.}~\bibnamefont {Martín-Jim\'enez}}, \bibinfo {author} {\bibfnamefont
  {B.}~\bibnamefont {Cirera}}, \bibinfo {author} {\bibfnamefont
  {F.}~\bibnamefont {Calleja}}, \bibinfo {author} {\bibfnamefont
  {S.}~\bibnamefont {D\'iaz-Tendero}}, \bibinfo {author} {\bibfnamefont
  {M.}~\bibnamefont {Alcam\'i}}, \bibinfo {author} {\bibfnamefont
  {L.}~\bibnamefont {Floreano}}, \bibinfo {author} {\bibfnamefont
  {M.}~\bibnamefont {Dom\'inguez-Rivera}}, \bibinfo {author} {\bibfnamefont
  {A.~L.}\ \bibnamefont {V\'azquez~de Parga}}, \bibinfo {author} {\bibfnamefont
  {D.}~\bibnamefont {\'Ecija}}, \bibinfo {author} {\bibfnamefont {J.~M.}\
  \bibnamefont {Gallego}}, \bibinfo {author} {\bibfnamefont {R.}~\bibnamefont
  {Miranda}}, \bibinfo {author} {\bibfnamefont {F.}~\bibnamefont {Martín}}, \
  and\ \bibinfo {author} {\bibfnamefont {R.}~\bibnamefont {Otero}},\ }\href
  {\doibase 10.1021/acs.jpcc.7b08017} {\bibfield  {journal} {\bibinfo
  {journal} {J. Phys. Chem. C}\ }\textbf {\bibinfo {volume} {121}},\ \bibinfo
  {pages} {23505} (\bibinfo {year} {2017})} \BibitemShut {NoStop}%
\bibitem [{\citenamefont {Fern{\'{a}}ndez-Torrente}\ \emph
  {et~al.}(2008)\citenamefont {Fern{\'{a}}ndez-Torrente}, \citenamefont
  {Franke},\ and\ \citenamefont {Pascual}}]{Fernandez-Torrente2008}%
  \BibitemOpen
  \bibfield  {author} {\bibinfo {author} {\bibfnamefont {I.}~\bibnamefont
  {Fern{\'{a}}ndez-Torrente}}, \bibinfo {author} {\bibfnamefont
  {K.}~\bibnamefont {Franke}}, \ and\ \bibinfo {author} {\bibfnamefont
  {J.}~\bibnamefont {Pascual}},\ }\href {\doibase
  10.1103/PhysRevLett.101.217203} {\bibfield  {journal} {\bibinfo  {journal}
  {Phys. Rev. Lett.}\ }\textbf {\bibinfo {volume} {101}},\ \bibinfo {pages}
  {217203} (\bibinfo {year} {2008})}\BibitemShut {NoStop}%
\bibitem [{\citenamefont {Fern\'andez-Torrente}\ \emph
  {et~al.}(2012)\citenamefont {Fern\'andez-Torrente}, \citenamefont
  {Kreikemeyer-Lorenzo}, \citenamefont {Str\'o\ifmmode~\dot{z}\else
  \.{z}\fi{}ecka}, \citenamefont {Franke},\ and\ \citenamefont
  {Pascual}}]{Torrente2012}%
  \BibitemOpen
  \bibfield  {author} {\bibinfo {author} {\bibfnamefont {I.}~\bibnamefont
  {Fern\'andez-Torrente}}, \bibinfo {author} {\bibfnamefont {D.}~\bibnamefont
  {Kreikemeyer-Lorenzo}}, \bibinfo {author} {\bibfnamefont {A.}~\bibnamefont
  {Str\'o\ifmmode~\dot{z}\else \.{z}\fi{}ecka}}, \bibinfo {author}
  {\bibfnamefont {K.~J.}\ \bibnamefont {Franke}}, \ and\ \bibinfo {author}
  {\bibfnamefont {J.~I.}\ \bibnamefont {Pascual}},\ }\href {\doibase
  10.1103/PhysRevLett.108.036801} {\bibfield  {journal} {\bibinfo  {journal}
  {Phys. Rev. Lett.}\ }\textbf {\bibinfo {volume} {108}},\ \bibinfo {pages}
  {036801} (\bibinfo {year} {2012})}\BibitemShut {NoStop}%
\bibitem [{\citenamefont {Umbach}\ \emph {et~al.}(2013)\citenamefont {Umbach},
  \citenamefont {Fern\'andez-Torrente}, \citenamefont {Ruby}, \citenamefont
  {Schulz}, \citenamefont {Lotze}, \citenamefont {Rurali}, \citenamefont
  {Persson}, \citenamefont {Pascual},\ and\ \citenamefont
  {Franke}}]{Umbach2013}%
  \BibitemOpen
  \bibfield  {author} {\bibinfo {author} {\bibfnamefont {T.~R.}\ \bibnamefont
  {Umbach}}, \bibinfo {author} {\bibfnamefont {I.}~\bibnamefont
  {Fern\'andez-Torrente}}, \bibinfo {author} {\bibfnamefont {M.}~\bibnamefont
  {Ruby}}, \bibinfo {author} {\bibfnamefont {F.}~\bibnamefont {Schulz}},
  \bibinfo {author} {\bibfnamefont {C.}~\bibnamefont {Lotze}}, \bibinfo
  {author} {\bibfnamefont {R.}~\bibnamefont {Rurali}}, \bibinfo {author}
  {\bibfnamefont {M.}~\bibnamefont {Persson}}, \bibinfo {author} {\bibfnamefont
  {J.~I.}\ \bibnamefont {Pascual}}, \ and\ \bibinfo {author} {\bibfnamefont
  {K.~J.}\ \bibnamefont {Franke}},\ }\href@noop {} {\bibfield  {journal}
  {\bibinfo  {journal} {New J. Phys.}\ }\textbf {\bibinfo {volume} {15}},\
  \bibinfo {pages} {083048} (\bibinfo {year} {2013})}\BibitemShut {NoStop}%
\bibitem [{\citenamefont {W\"ackerlin}\ \emph {et~al.}(2011)\citenamefont
  {W\"ackerlin}, \citenamefont {Iacovita}, \citenamefont {Chylarecka},
  \citenamefont {Fesser}, \citenamefont {Jung},\ and\ \citenamefont
  {Ballav}}]{Wackerlin2011}%
  \BibitemOpen
  \bibfield  {author} {\bibinfo {author} {\bibfnamefont {C.}~\bibnamefont
  {W\"ackerlin}}, \bibinfo {author} {\bibfnamefont {C.}~\bibnamefont
  {Iacovita}}, \bibinfo {author} {\bibfnamefont {D.}~\bibnamefont
  {Chylarecka}}, \bibinfo {author} {\bibfnamefont {P.}~\bibnamefont {Fesser}},
  \bibinfo {author} {\bibfnamefont {T.~A.}\ \bibnamefont {Jung}}, \ and\
  \bibinfo {author} {\bibfnamefont {N.}~\bibnamefont {Ballav}},\ }\href
  {\doibase 10.1039/C1CC12519B} {\bibfield  {journal} {\bibinfo  {journal}
  {Chem. Commun.}\ }\textbf {\bibinfo {volume} {47}},\ \bibinfo {pages} {9146}
  (\bibinfo {year} {2011})}\BibitemShut {NoStop}%
\bibitem [{\citenamefont {Zhao}\ \emph {et~al.}(2005)\citenamefont {Zhao},
  \citenamefont {Li}, \citenamefont {Chen}, \citenamefont {Xiang},
  \citenamefont {Wang}, \citenamefont {Pan}, \citenamefont {Wang},
  \citenamefont {Xiao}, \citenamefont {Yang}, \citenamefont {Hou},\ and\
  \citenamefont {Zhu}}]{Zhao2005}%
  \BibitemOpen
  \bibfield  {author} {\bibinfo {author} {\bibfnamefont {A.}~\bibnamefont
  {Zhao}}, \bibinfo {author} {\bibfnamefont {Q.}~\bibnamefont {Li}}, \bibinfo
  {author} {\bibfnamefont {L.}~\bibnamefont {Chen}}, \bibinfo {author}
  {\bibfnamefont {H.}~\bibnamefont {Xiang}}, \bibinfo {author} {\bibfnamefont
  {W.}~\bibnamefont {Wang}}, \bibinfo {author} {\bibfnamefont {S.}~\bibnamefont
  {Pan}}, \bibinfo {author} {\bibfnamefont {B.}~\bibnamefont {Wang}}, \bibinfo
  {author} {\bibfnamefont {X.}~\bibnamefont {Xiao}}, \bibinfo {author}
  {\bibfnamefont {J.}~\bibnamefont {Yang}}, \bibinfo {author} {\bibfnamefont
  {J.~G.}\ \bibnamefont {Hou}}, \ and\ \bibinfo {author} {\bibfnamefont
  {Q.}~\bibnamefont {Zhu}},\ }\href {\doibase 10.1126/science.1113449}
  {\bibfield  {journal} {\bibinfo  {journal} {Science}\ }\textbf {\bibinfo
  {volume} {309}},\ \bibinfo {pages} {1542} (\bibinfo {year} {2005})}\BibitemShut
  {NoStop}%
\bibitem [{\citenamefont {Iancu}\ \emph {et~al.}(2006)\citenamefont {Iancu},
  \citenamefont {Deshpande},\ and\ \citenamefont {Hla}}]{Iancu2006}%
  \BibitemOpen
  \bibfield  {author} {\bibinfo {author} {\bibfnamefont {V.}~\bibnamefont
  {Iancu}}, \bibinfo {author} {\bibfnamefont {A.}~\bibnamefont {Deshpande}}, \
  and\ \bibinfo {author} {\bibfnamefont {S.-W.}\ \bibnamefont {Hla}},\ }\href
  {\doibase 10.1021/nl0601886} {\bibfield  {journal} {\bibinfo  {journal} {Nano
  Lett.}\ }\textbf {\bibinfo {volume} {6}},\ \bibinfo {pages} {820} (\bibinfo
  {year} {2006})} \BibitemShut {NoStop}%
\bibitem [{\citenamefont {Fernandez-Torrente}\ \emph
  {et~al.}(2007)\citenamefont {Fernandez-Torrente}, \citenamefont {Monturet},
  \citenamefont {Franke}, \citenamefont {Fraxedas}, \citenamefont {Lorente},\
  and\ \citenamefont {Pascual}}]{Fernandez-Torrente2007}%
  \BibitemOpen
  \bibfield  {author} {\bibinfo {author} {\bibfnamefont {I.}~\bibnamefont
  {Fernandez-Torrente}}, \bibinfo {author} {\bibfnamefont {S.}~\bibnamefont
  {Monturet}}, \bibinfo {author} {\bibfnamefont {K.}~\bibnamefont {Franke}},
  \bibinfo {author} {\bibfnamefont {J.}~\bibnamefont {Fraxedas}}, \bibinfo
  {author} {\bibfnamefont {N.}~\bibnamefont {Lorente}}, \ and\ \bibinfo
  {author} {\bibfnamefont {J.}~\bibnamefont {Pascual}},\ }\href {\doibase
  10.1103/PhysRevLett.99.176103} {\bibfield  {journal} {\bibinfo  {journal}
  {Phys. Rev. Lett.}\ }\textbf {\bibinfo {volume} {99}},\ \bibinfo {pages}
  {176103} (\bibinfo {year} {2007})}\BibitemShut {NoStop}%
\bibitem [{\citenamefont {Ternes}\ \emph {et~al.}(2009)\citenamefont {Ternes},
  \citenamefont {Heinrich},\ and\ \citenamefont {Schneider}}]{Ternes2009}%
  \BibitemOpen
  \bibfield  {author} {\bibinfo {author} {\bibfnamefont {M.}~\bibnamefont
  {Ternes}}, \bibinfo {author} {\bibfnamefont {A.~J.}\ \bibnamefont
  {Heinrich}}, \ and\ \bibinfo {author} {\bibfnamefont {W.-D.}\ \bibnamefont
  {Schneider}},\ }\href {\doibase 10.1088/0953-8984/21/5/053001} {\bibfield
  {journal} {\bibinfo  {journal} {J. Phys.: Cond. Matt.}\ }\textbf {\bibinfo
  {volume} {21}},\ \bibinfo {pages} {53001} (\bibinfo {year}
  {2009})}\BibitemShut {NoStop}%
\bibitem [{\citenamefont {Stipe}(1998)}]{Stipe1998a}%
  \BibitemOpen
  \bibfield  {author} {\bibinfo {author} {\bibfnamefont {B.~C.}\ \bibnamefont
  {Stipe}}, \bibinfo {author} {\bibfnamefont {M.~A.}\ \bibnamefont
  {Rezaei}}, \ and\ \bibinfo {author} {\bibfnamefont {W.}\ \bibnamefont
  {Ho}}\ }\href {\doibase 10.1126/science.280.5370.1732} {\bibfield
  {journal} {\bibinfo  {journal} {Science}\ }\textbf {\bibinfo {volume}
  {280}},\ \bibinfo {pages} {1732} (\bibinfo {year} {1998})}\BibitemShut
  {NoStop}%
\bibitem [{\citenamefont {Wegner}\ \emph {et~al.}(2009)\citenamefont {Wegner},
  \citenamefont {Yamachika}, \citenamefont {Zhang}, \citenamefont {Wang},
  \citenamefont {Baruah}, \citenamefont {Pederson}, \citenamefont {Bartlett},
  \citenamefont {Long},\ and\ \citenamefont {Crommie}}]{Wegner2009}%
  \BibitemOpen
  \bibfield  {author} {\bibinfo {author} {\bibfnamefont {D.}~\bibnamefont
  {Wegner}}, \bibinfo {author} {\bibfnamefont {R.}~\bibnamefont {Yamachika}},
  \bibinfo {author} {\bibfnamefont {X.}~\bibnamefont {Zhang}}, \bibinfo
  {author} {\bibfnamefont {Y.}~\bibnamefont {Wang}}, \bibinfo {author}
  {\bibfnamefont {T.}~\bibnamefont {Baruah}}, \bibinfo {author} {\bibfnamefont
  {M.}~\bibnamefont {Pederson}}, \bibinfo {author} {\bibfnamefont
  {B.}~\bibnamefont {Bartlett}}, \bibinfo {author} {\bibfnamefont
  {J.}~\bibnamefont {Long}}, \ and\ \bibinfo {author} {\bibfnamefont
  {M.}~\bibnamefont {Crommie}},\ }\href {\doibase
  10.1103/PhysRevLett.103.087205} {\bibfield  {journal} {\bibinfo  {journal}
  {Phys. Rev. Lett.}\ }\textbf {\bibinfo {volume} {103}},\ \bibinfo {pages}
  {87205} (\bibinfo {year} {2009})}\BibitemShut {NoStop}%
\bibitem [{\citenamefont {Michaelian}\ \emph {et~al.}(1982)\citenamefont
  {Michaelian}, \citenamefont {Rieckhoff},\ and\ \citenamefont
  {Voigt}}]{Michaelian1982}%
  \BibitemOpen
  \bibfield  {author} {\bibinfo {author} {\bibfnamefont {K.}~\bibnamefont
  {Michaelian}}, \bibinfo {author} {\bibfnamefont {K.}~\bibnamefont
  {Rieckhoff}}, \ and\ \bibinfo {author} {\bibfnamefont {E.}~\bibnamefont
  {Voigt}},\ }\href {\doibase https://doi.org/10.1016/0022-2852(82)90230-2}
  {\bibfield  {journal} {\bibinfo  {journal} {J. Mol. Spec.}\ }\textbf
  {\bibinfo {volume} {95}},\ \bibinfo {pages} {1 } (\bibinfo {year}
  {1982})}\BibitemShut {NoStop}%
\bibitem [{\citenamefont {Wegner}\ \emph {et~al.}(2013)\citenamefont {Wegner},
  \citenamefont {Yamachika}, \citenamefont {Zhang}, \citenamefont {Wang},
  \citenamefont {Crommie},\ and\ \citenamefont {Lorente}}]{Wegner2013}%
  \BibitemOpen
  \bibfield  {author} {\bibinfo {author} {\bibfnamefont {D.}~\bibnamefont
  {Wegner}}, \bibinfo {author} {\bibfnamefont {R.}~\bibnamefont {Yamachika}},
  \bibinfo {author} {\bibfnamefont {X.}~\bibnamefont {Zhang}}, \bibinfo
  {author} {\bibfnamefont {Y.}~\bibnamefont {Wang}}, \bibinfo {author}
  {\bibfnamefont {M.~F.}\ \bibnamefont {Crommie}}, \ and\ \bibinfo {author}
  {\bibfnamefont {N.}~\bibnamefont {Lorente}},\ }\href {\doibase
  10.1021/nl304081q} {\bibfield  {journal} {\bibinfo  {journal} {Nano Lett.}\
  }\textbf {\bibinfo {volume} {13}},\ \bibinfo {pages} {2346} (\bibinfo {year}
  {2013})} \BibitemShut {NoStop}%
\bibitem [{\citenamefont {Paaske}\ and\ \citenamefont
  {Flensberg}(2005)}]{Paaske2005}%
  \BibitemOpen
  \bibfield  {author} {\bibinfo {author} {\bibfnamefont {J.}~\bibnamefont
  {Paaske}}\ and\ \bibinfo {author} {\bibfnamefont {K.}~\bibnamefont
  {Flensberg}},\ }\href {\doibase 10.1103/PhysRevLett.94.176801} {\bibfield
  {journal} {\bibinfo  {journal} {Phys. Rev. Lett.}\ }\textbf {\bibinfo
  {volume} {94}},\ \bibinfo {pages} {1} (\bibinfo {year} {2005})}\BibitemShut
  {NoStop}%
\bibitem [{\citenamefont {Iancu}\ \emph {et~al.}(2016)\citenamefont {Iancu},
  \citenamefont {Schouteden}, \citenamefont {Li},\ and\ \citenamefont
  {Van~Haesendonck}}]{Iancu2016}%
  \BibitemOpen
  \bibfield  {author} {\bibinfo {author} {\bibfnamefont {V.}~\bibnamefont
  {Iancu}}, \bibinfo {author} {\bibfnamefont {K.}~\bibnamefont {Schouteden}},
  \bibinfo {author} {\bibfnamefont {Z.}~\bibnamefont {Li}}, \ and\ \bibinfo
  {author} {\bibfnamefont {C.}~\bibnamefont {Van~Haesendonck}},\ }\href
  {\doibase 10.1039/C6CC03847F} {\bibfield  {journal} {\bibinfo  {journal}
  {Chem. Commun.}\ }\textbf {\bibinfo {volume} {52}},\ \bibinfo {pages} {11359}
  (\bibinfo {year} {2016})}\BibitemShut {NoStop}%
\bibitem [{\citenamefont {Wegner}\ \emph {et~al.}(2008)\citenamefont {Wegner},
  \citenamefont {Yamachika}, \citenamefont {Wang}, \citenamefont {Brar},
  \citenamefont {Bartlett}, \citenamefont {Long},\ and\ \citenamefont
  {Crommie}}]{Wegner2008}%
  \BibitemOpen
  \bibfield  {author} {\bibinfo {author} {\bibfnamefont {D.}~\bibnamefont
  {Wegner}}, \bibinfo {author} {\bibfnamefont {R.}~\bibnamefont {Yamachika}},
  \bibinfo {author} {\bibfnamefont {Y.}~\bibnamefont {Wang}}, \bibinfo {author}
  {\bibfnamefont {V.~W.}\ \bibnamefont {Brar}}, \bibinfo {author}
  {\bibfnamefont {B.~M.}\ \bibnamefont {Bartlett}}, \bibinfo {author}
  {\bibfnamefont {J.~R.}\ \bibnamefont {Long}}, \ and\ \bibinfo {author}
  {\bibfnamefont {M.~F.}\ \bibnamefont {Crommie}},\ }\href {\doibase
  10.1021/nl072217y} {\bibfield  {journal} {\bibinfo  {journal} {Nano Lett.}\
  }\textbf {\bibinfo {volume} {8}},\ \bibinfo {pages} {131} (\bibinfo {year}
  {2008})}\BibitemShut {NoStop}%
\bibitem [{\citenamefont {Gross}\ \emph
  {et~al.}(2009{\natexlab{a}})\citenamefont {Gross}, \citenamefont {Mohn},
  \citenamefont {Liljeroth}, \citenamefont {Repp}, \citenamefont {Giessibl},\
  and\ \citenamefont {Meyer}}]{Gross2009a}%
  \BibitemOpen
  \bibfield  {author} {\bibinfo {author} {\bibfnamefont {L.}~\bibnamefont
  {Gross}}, \bibinfo {author} {\bibfnamefont {F.}~\bibnamefont {Mohn}},
  \bibinfo {author} {\bibfnamefont {P.}~\bibnamefont {Liljeroth}}, \bibinfo
  {author} {\bibfnamefont {J.}~\bibnamefont {Repp}}, \bibinfo {author}
  {\bibfnamefont {F.~J.}\ \bibnamefont {Giessibl}}, \ and\ \bibinfo {author}
  {\bibfnamefont {G.}~\bibnamefont {Meyer}},\ }\href {\doibase
  10.1126/science.1172273} {\bibfield  {journal} {\bibinfo  {journal}
  {Science}\ }\textbf {\bibinfo {volume} {324}},\ \bibinfo {pages} {1428}
  (\bibinfo {year} {2009}{\natexlab{a}})} \BibitemShut
  {NoStop}%
\bibitem [{\citenamefont {Mohn}\ \emph {et~al.}(2002)\citenamefont {Mohn},
  \citenamefont {Gross}, \citenamefont {Moll},\ and\ \citenamefont
  {Meyer}}]{Mohn2012}%
  \BibitemOpen
  \bibfield  {author} {\bibinfo {author} {\bibfnamefont {F.}~\bibnamefont
  {Mohn}}, \bibinfo {author} {\bibfnamefont {L.}~\bibnamefont {Gross}},
  \bibinfo {author} {\bibfnamefont {N.}~\bibnamefont {Moll}}, \ and\ \bibinfo
  {author} {\bibfnamefont {G.}~\bibnamefont {Meyer}},\ }\href {\doibase
  10.1038/nnano.2012.20} {\bibfield  {journal} {\bibinfo  {journal} {Nature
  Nanotech.}\ }\textbf {\bibinfo {volume} {7}},\ \bibinfo {pages} {227}
  (\bibinfo {year} {2002})}\BibitemShut {NoStop}%
\bibitem [{\citenamefont {Gross}\ \emph
  {et~al.}(2009{\natexlab{b}})\citenamefont {Gross}, \citenamefont {Mohn},
  \citenamefont {Moll}, \citenamefont {Liljeroth},\ and\ \citenamefont
  {Meyer}}]{Gross2009b}%
  \BibitemOpen
  \bibfield  {author} {\bibinfo {author} {\bibfnamefont {L.}~\bibnamefont
  {Gross}}, \bibinfo {author} {\bibfnamefont {F.}~\bibnamefont {Mohn}},
  \bibinfo {author} {\bibfnamefont {N.}~\bibnamefont {Moll}}, \bibinfo {author}
  {\bibfnamefont {P.}~\bibnamefont {Liljeroth}}, \ and\ \bibinfo {author}
  {\bibfnamefont {G.}~\bibnamefont {Meyer}},\ }\href {\doibase
  10.1126/science.1176210} {\bibfield  {journal} {\bibinfo  {journal}
  {Science}\ }\textbf {\bibinfo {volume} {325}},\ \bibinfo {pages} {1110}
  (\bibinfo {year} {2009}{\natexlab{b}})}\BibitemShut {NoStop}%
\bibitem [{Note1()}]{Note1}%
  \BibitemOpen
  \bibinfo {note} {$df$ data was corrected for $Z$-creep by setting the initial
  and final point of the forward-backward spectrum to the same value. The
  deviation throughout the forward-backward scan was fitted by a linear and one
  exponentially decaying $Z$ value as correction factor. After this correction
  the largest error stems from the noise in the $df$ signal and the fact that
  the stable $df-V$ range is rather small.}\BibitemShut {Stop}%
\bibitem [{\citenamefont {Schuler}\ \emph {et~al.}(2014)\citenamefont
  {Schuler}, \citenamefont {Liu}, \citenamefont {Geng}, \citenamefont
  {Decurtins}, \citenamefont {Meyer},\ and\ \citenamefont
  {Gross}}]{Schuler2014}%
  \BibitemOpen
  \bibfield  {author} {\bibinfo {author} {\bibfnamefont {B.}~\bibnamefont
  {Schuler}}, \bibinfo {author} {\bibfnamefont {S.-X.}\ \bibnamefont {Liu}},
  \bibinfo {author} {\bibfnamefont {Y.}~\bibnamefont {Geng}}, \bibinfo {author}
  {\bibfnamefont {S.}~\bibnamefont {Decurtins}}, \bibinfo {author}
  {\bibfnamefont {G.}~\bibnamefont {Meyer}}, \ and\ \bibinfo {author}
  {\bibfnamefont {L.}~\bibnamefont {Gross}},\ }\href {\doibase
  10.1021/nl500805x} {\bibfield  {journal} {\bibinfo  {journal} {Nano Lett.}\
  }\textbf {\bibinfo {volume} {14}},\ \bibinfo {pages} {3342} (\bibinfo {year}
  {2014})} \BibitemShut {NoStop}%
\bibitem [{\citenamefont {Moll}\ \emph {et~al.}(2010)\citenamefont {Moll},
  \citenamefont {Gross}, \citenamefont {Mohn}, \citenamefont {Curioni},\ and\
  \citenamefont {Meyer}}]{Moll2010}%
  \BibitemOpen
  \bibfield  {author} {\bibinfo {author} {\bibfnamefont {N.}~\bibnamefont
  {Moll}}, \bibinfo {author} {\bibfnamefont {L.}~\bibnamefont {Gross}},
  \bibinfo {author} {\bibfnamefont {F.}~\bibnamefont {Mohn}}, \bibinfo {author}
  {\bibfnamefont {A.}~\bibnamefont {Curioni}}, \ and\ \bibinfo {author}
  {\bibfnamefont {G.}~\bibnamefont {Meyer}},\ }\href {\doibase
  10.1088/1367-2630/12/12/125020} {\bibfield  {journal} {\bibinfo  {journal}
  {New J. Phys.}\ }\textbf {\bibinfo {volume} {12}},\ \bibinfo {pages} {125020}
  (\bibinfo {year} {2010})}\BibitemShut {NoStop}%
\bibitem [{\citenamefont {{Nano Surf Group}}()}]{HapalaCode}%
  \BibitemOpen
  \bibfield  {author} {\bibinfo {author} {\bibnamefont {{Nano Surf Group}}},\
  }\href {http://nanosurf.fzu.cz/wiki/doku.php?id=probe_particle_model}
  {\enquote {\bibinfo {title} {Probe particle model},}\ }\BibitemShut {NoStop}%
\bibitem [{\citenamefont {Hapala}\ \emph
  {et~al.}(2014{\natexlab{a}})\citenamefont {Hapala}, \citenamefont {Kichin},
  \citenamefont {Wagner}, \citenamefont {Tautz}, \citenamefont {Temirov},\ and\
  \citenamefont {Jel{\'{i}}nek}}]{Hapala2014}%
  \BibitemOpen
  \bibfield  {author} {\bibinfo {author} {\bibfnamefont {P.}~\bibnamefont
  {Hapala}}, \bibinfo {author} {\bibfnamefont {G.}~\bibnamefont {Kichin}},
  \bibinfo {author} {\bibfnamefont {C.}~\bibnamefont {Wagner}}, \bibinfo
  {author} {\bibfnamefont {F.~S.}\ \bibnamefont {Tautz}}, \bibinfo {author}
  {\bibfnamefont {R.}~\bibnamefont {Temirov}}, \ and\ \bibinfo {author}
  {\bibfnamefont {P.}~\bibnamefont {Jel{\'{i}}nek}},\ }\href {\doibase
  10.1103/PhysRevB.90.085421} {\bibfield  {journal} {\bibinfo  {journal} {Phys.
  Rev. B}\ }\textbf {\bibinfo {volume} {90}},\ \bibinfo {pages} {085421}
  (\bibinfo {year} {2014}{\natexlab{a}})}\BibitemShut {NoStop}%
\bibitem [{\citenamefont {Hapala}\ \emph
  {et~al.}(2014{\natexlab{b}})\citenamefont {Hapala}, \citenamefont {Temirov},
  \citenamefont {Tautz},\ and\ \citenamefont {Jel{\'{i}}nek}}]{Hapala2014b}%
  \BibitemOpen
  \bibfield  {author} {\bibinfo {author} {\bibfnamefont {P.}~\bibnamefont
  {Hapala}}, \bibinfo {author} {\bibfnamefont {R.}~\bibnamefont {Temirov}},
  \bibinfo {author} {\bibfnamefont {F.~S.}\ \bibnamefont {Tautz}}, \ and\
  \bibinfo {author} {\bibfnamefont {P.}~\bibnamefont {Jel{\'{i}}nek}},\ }\href
  {\doibase 10.1103/PhysRevLett.113.226101} {\bibfield  {journal} {\bibinfo
  {journal} {Phys. Rev. Lett.}\ }\textbf {\bibinfo {volume} {113}},\ \bibinfo
  {pages} {226101} (\bibinfo {year} {2014}{\natexlab{b}})}\BibitemShut
  {NoStop}%
\bibitem [{\citenamefont {Kortright}\ \emph {et~al.}(2008)\citenamefont
  {Kortright}, \citenamefont {Lincoln}, \citenamefont {Edelstein},\ and\
  \citenamefont {Epstein}}]{Kortright2008}%
  \BibitemOpen
  \bibfield  {author} {\bibinfo {author} {\bibfnamefont {J.~B.}\ \bibnamefont
  {Kortright}}, \bibinfo {author} {\bibfnamefont {D.~M.}\ \bibnamefont
  {Lincoln}}, \bibinfo {author} {\bibfnamefont {R.~S.}\ \bibnamefont
  {Edelstein}}, \ and\ \bibinfo {author} {\bibfnamefont {A.~J.}\ \bibnamefont
  {Epstein}},\ }\href {\doibase 10.1103/PhysRevLett.100.257204} {\bibfield
  {journal} {\bibinfo  {journal} {Phys. Rev. Lett.}\ }\textbf {\bibinfo
  {volume} {100}},\ \bibinfo {pages} {257204} (\bibinfo {year}
  {2008})}\BibitemShut {NoStop}%
\bibitem [{\citenamefont {Choi}\ \emph {et~al.}(2010)\citenamefont {Choi},
  \citenamefont {Bedwani}, \citenamefont {Rochefort}, \citenamefont {Chen},
  \citenamefont {Epstein},\ and\ \citenamefont {Gupta}}]{Choi2010}%
  \BibitemOpen
  \bibfield  {author} {\bibinfo {author} {\bibfnamefont {T.}~\bibnamefont
  {Choi}}, \bibinfo {author} {\bibfnamefont {S.}~\bibnamefont {Bedwani}},
  \bibinfo {author} {\bibfnamefont {A.}~\bibnamefont {Rochefort}}, \bibinfo
  {author} {\bibfnamefont {C.-Y.}\ \bibnamefont {Chen}}, \bibinfo {author}
  {\bibfnamefont {A.~J.}\ \bibnamefont {Epstein}}, \ and\ \bibinfo {author}
  {\bibfnamefont {J.}~\bibnamefont {Gupta}},\ }\href {\doibase
  10.1021/nl1024563} {\bibfield  {journal} {\bibinfo  {journal} {Nano Lett.}\
  }\textbf {\bibinfo {volume} {10}},\ \bibinfo {pages} {4175} (\bibinfo {year}
  {2010})}\BibitemShut {NoStop}%
\bibitem [{\citenamefont {Miessler}\ and\ \citenamefont
  {Tarr}(2011)}]{miessler2011inorganic}%
  \BibitemOpen
  \bibfield  {author} {\bibinfo {author} {\bibfnamefont {G.}~\bibnamefont
  {Miessler}}\ and\ \bibinfo {author} {\bibfnamefont {D.}~\bibnamefont
  {Tarr}},\ }\href {https://books.google.de/books?id=c-QEQwAACAAJ} {\emph
  {\bibinfo {title} {Inorganic Chemistry}}}\ (\bibinfo  {publisher} {Pearson
  Prentice Hall},\ \bibinfo {year} {2011})\BibitemShut {NoStop}%
\end{thebibliography}

%

\end{document}